\documentclass[letterpaper]{article} 
\usepackage{aaai25}  
\usepackage{times}  
\usepackage{helvet}  
\usepackage{courier}  
\usepackage[hyphens]{url}  
\usepackage{graphicx} 
\usepackage{natbib}  
\usepackage{caption} 
\usepackage{algorithm}
\usepackage{algorithmic}
\usepackage{multirow}
\usepackage{epsfig}
\usepackage{amsmath}
\usepackage{amssymb}
\usepackage{comment}
\usepackage{color}
\usepackage{cite}
\usepackage{xcolor}
\usepackage{amsmath}
\usepackage{multirow}
\usepackage{graphicx}
\usepackage{subfig}
\usepackage{enumitem}
\usepackage{adjustbox}
\usepackage{color, colortbl}
\definecolor{LightCyan}{rgb}{0.92,0.92,1}
\definecolor{Gray}{gray}{0.9}
\usepackage{hhline}
\usepackage{makecell}
\nocopyright
\usepackage{newfloat}
\usepackage{listings}
\DeclareCaptionStyle{ruled}{labelfont=normalfont,labelsep=colon,strut=off} 
\floatstyle{ruled}
\newfloat{listing}{tb}{lst}{}
\floatname{listing}{Listing}
\title{EmoReg: Directional Latent Vector Modeling for Emotional Intensity Regularization in Diffusion-based Voice Conversion}
\author {
    Ashishkumar Gudmalwar\textsuperscript{\rm 1},
    Ishan D. Biyani\textsuperscript{\rm 1},
    Nirmesh Shah\textsuperscript{\rm 1},
    Pankaj Wasnik\textsuperscript{\rm 1},
    Rajiv Ratn Shah\textsuperscript{\rm 2}
}
\affiliations {
    \textsuperscript{\rm 1}Media Analysis Group, Sony Research India, Bangalore\\
    \textsuperscript{\rm 2}Indraprastha Institute of Information Technology (IIIT), Delhi, India\\
    \{ashish.gudmalwar1,ishan.biyani,nirmesh.shah,pankaj.wasnik\}@sony.com, rajivratn@iiitd.ac.in
}

\usepackage{bibentry}
\begin{document}

\maketitle

\begin{abstract} 
\label{sec:abstract}
The Emotional Voice Conversion (EVC) aims to convert the discrete emotional state from the source emotion to the target for a given speech utterance while preserving linguistic content. In this paper, we propose regularizing emotion intensity in the diffusion-based EVC framework to generate precise speech of the target emotion. Traditional approaches control the intensity of an emotional state in the utterance via emotion class probabilities or intensity labels that often lead to inept style manipulations and degradations in quality. On the contrary, we aim to regulate emotion intensity using self-supervised learning-based feature representations and unsupervised directional latent vector modeling (DVM) in the emotional embedding space within a diffusion-based framework.
These emotion embeddings can be modified based on the given target emotion intensity and the corresponding direction vector. Furthermore, the updated embeddings can be fused in the reverse diffusion process to generate the speech with the desired emotion and intensity. In summary, this paper aims to achieve high-quality emotional intensity regularization in the diffusion-based EVC framework, which is the first of its kind work. The effectiveness of the proposed method has been shown across state-of-the-art (SOTA) baselines in terms of subjective and objective evaluations for the English and Hindi languages\footnote{Demo samples are available at the following URL: \\
\url{https://nirmesh-sony.github.io/EmoReg/}}. 
\end{abstract}

%

\section{Introduction}
\label{sec:Introduction}
Despite significant progress in the field of Generative AI, speech synthesis models still encounter several challenges when it comes to the AI-based dubbing of entertainment content such as movies and serials \cite{brannon2023dubbing,wu2023videodubber,hu2021neural,mhaskar-etal-2024-isometric,sahipjohn2024dubwise}. AI-based dubbing involves replicating input speech emotion and controlling its intensity depending on the context and emotion of the scene \cite{brannon2023dubbing,zhou2022emotion,zhou2022mixed,amiriparian2023guest}. Most of today's text-to-speech (TTS) systems can produce high-quality, high-fidelity, natural speech output, but they still lack expressiveness and fine control over emotional states \cite{barakat2024deep,gudmalwar2024vecl}. Hence, professional voice-over/dubbing artists are still preferred in the dubbing industry to tackle the complex demands of generating emotionally engaging speech \cite{gutentag2017successful}. This poses challenges in terms of dubbing at a large scale, turnaround time, and operational costs.

One of the major roadblocks in building emotional speech synthesis is the unavailability of a large emotional speech database with diverse emotional expressions and a wide range of emotion intensities. Even annotating accurate emotional states in the speech requires expert knowledge and is costly and labor-intensive as labeling each speech file with the correct emotion and intensity involves detailed subjective assessments. Hence, emotional intensity control or regularization is still an under-explored open research problem. To this end, we primarily focus on the emotional voice conversion (EVC) task, a sub-topic of emotional speech synthesis for emotional intensity regularization \cite{zhou2022emotion,zhou2021seen,zhou2022mixed,shah2023nonparallel}. In particular, we employ a self-supervised learning (SSL)-based framework to tackle the issues related to the unavailability of a large annotated emotional speech database. 
\begin{figure}[t]
    \centering    
    \includegraphics[width=\linewidth]{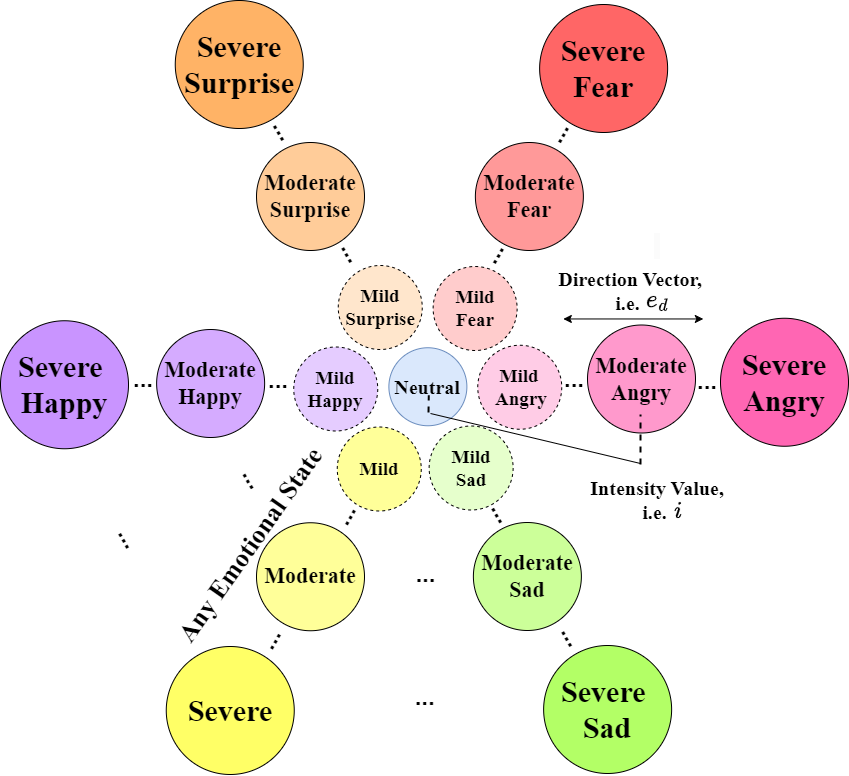}
    \caption{Conceptual representation of emotional intensity regularization based on direction vector and intensity value.}
    \label{fig:intro_emoreg}
\end{figure}

\indent The EVC aims to alter the discrete emotional state of a speech utterance while preserving linguistic content and the speaker's identity \cite{zhou2021seen}. Furthermore, emotional intensity regularization involves fine control over intensity associated with the target emotional state. For example, a neutral utterance can be converted into a mild, moderate, or severe level of anger emotion, as shown in Figure \ref{fig:intro_emoreg}. Current EVC methods usually tackle this scenario with discrete \cite{zhou2022esd,shah2023nonparallel,zhou2022emotion,ekman1971constants} or continuous emotion representations \cite{russell1980circumplex,posner2005circumplex,cho2024emosphere}. Discrete emotion representations contain categorical labels such as happy, angry, sad, etc. On the other hand, continuous emotion representations are obtained from the circumplex models and contain continuous-independent dimensions, such as arousal and valance \cite{russell1980circumplex,posner2005circumplex}. Achieving emotion intensity control in continuous emotion representation is relatively easier than discrete emotion representations \cite{gunes2011emotion,prabhu2023wild,zhou2022emotion,zhou2022mixed,cho2024emosphere}. However, obtaining such continuous emotion representations-based annotated data is challenging, as discussed earlier. Hence, this paper aims to achieve emotion intensity regularization for the discrete emotion representations-based EVC methods. 

In summary, we propose a method to achieve emotion intensity regularization by means of self-supervised learning and direction latent vector modeling using a diffusion-based EVC. Our key contributions can be summarized as follows:
\begin{itemize}
    \item To the best of the authors` knowledge this is the first attempt that achieves a high-quality emotional intensity regularization in the diffusion-based EVC framework.
    \item We propose a novel direction latent vector modeling-based approach for obtaining fine control over intensity while transitioning across different emotional states. 
    \item The proposed EmoReg utilizes the SSL-based audio feature representations, which are obtained after finetuning the SSL-based framework for a downstream task related to emotions classification.
    \item Effectiveness of the proposed EmoReg approach has been shown against SOTA baselines with relevant subjective and objective evaluation techniques across two languages, namely English and Hindi.
\end{itemize}
\section{Related Work}
\indent EVC methods are usually categorized into parallel and non-parallel approaches depending on the nature of the training data. Parallel means each speaker has spoken the same utterance in different emotional states and non-parallel means recorded sentences are different depending on the emotional states. Since emotion is also dependent on the content that is being spoken, hence, researchers have primarily focused on non-parallel EVC approaches \cite{shah2023nonparallel}. Traditionally, researchers have explored various generative models for the discrete emotion representation-based EVC tasks, namely, GMM \cite{aihara2012gmm}, HMM \cite{inanoglu2007system}, DNN \cite{lorenzo2018investigating}, Sequence-to-sequence models \cite{ming2016deep}, Variational Auto Encoder (VAE) \cite{zhou2021vaw}, Generative Adversarial Networks (GAN) and its variants \cite{shah2023nonparallel,li2021starganv2}, Diffusion Model-based approaches \cite{prabhu2024emoconv}, etc. However, none of these approaches tackles scenarios of emotion intensity regularization. 

\indent Broadly, emotional intensity regularization is achieved from the emotional labels and the reference speech utterance. The emotion label-based approach is utilized to manipulate auxiliary features such as attention weights, saliency maps, ranking function, or signal processing attributes to achieve emotion intensity regularization \cite{zhou2022emotion,zhou2022mixed,matsumoto2020controlling,schnell2021improving,choi2021sequence,um2020emotional}. Among these, one of the most prominent techniques is relative attribute \cite{zhu2019controlling,zhou2022emotion,zhou2022mixed}, which describes the distinctions between binary classes using a learned ranking function. However, these methods fail to capture high-level complex abstract representations, which results in artifacts in the converted output and leads to significant degradation in the quality. Also, emotional manipulations often involve inter-dependencies across different types of features, hence manipulating any single feature in isolation fails in achieving emotional intensity regularization. On the other hand, some simple approaches manipulate learned emotion representations via scaling \cite{choi2021sequence} or interpolations \cite{um2020emotional} to achieve emotion regularization. However, these approaches do not work well since emotional embedding space does not align well with the assumption of linear interpolation. Furthermore, high-dimensional emotional embedding space is often sparsely populated between two styles. Thus, such approaches result in inept style manipulations.

\indent Another emotional intensity control-based strategy leverages the extended emotional dimensions, namely, Arousal, Valence, and Dominance (AVD) \cite{russell1980circumplex}.  AVD-based emotional dimensions provide continuous and fine-grained emotion representations, which can be utilized to alter emotional states in a continuous manner more appropriately than discrete human emotions-based representations. However, obtaining such annotations is difficult due to inherent subjectivity-related issues associated with different annotators and the high costs associated with preparing such data. Hence, emotional intensity regularization is an under-explored research area. 

\indent In summary, all previous discrete and continuous emotional intensity control methods produce low-quality noisy converted output. To tackle quality-related issues recently, diffusion-based models have seen great success as a potent and versatile generative AI technology in computer vision, audio, and reinforcement learning \cite{ho2020denoising}. Diffusion models serve as samplers in these applications, producing new samples while actively guiding them toward task-desired features  \cite{ho2020denoising}. They also offer versatile high-dimensional data modeling  \cite{ho2020denoising}. 
The diffusion model-based approach has also obtained remarkable success in the voice conversion \cite{popov2021diffusion} and EVC task \cite{prabhu2024emoconv}. However, it has not been explored for achieving emotion intensity control-related tasks. Hence, we exploit the diffusion-based EVC method with the proposed DVM-based strategy for the emotion intensity regularization task. The next section presents details about the proposed diffusion-based EmoReg architecture.

\section{Proposed Methodology}
\label{sec:ProposedApproach}
\subsection{Problem Formulation}
Given input speech $x_0$ along with reference emotion speech $x_{r}$ and intensity value $i$, the goal of EVC is to generate emotional intensity regularized converted speech $y$ such that $Y$ = $G(X_{0}, e_{s}, e_{r}, i, \theta)$, where $X_{0}$ and $Y$ denotes Mel-spectrogram features corresponding to the input speech $x_0$ and the converted emotional speech $y$, respectively. In addition, $e_{s}$, $e_{r}$ denotes emotional feature representation for source and reference speech, respectively. $\theta$ is model parameters corresponding to the considered conditional diffusion probabilistic model-based EVC architecture. Details of the proposed architecture is as follows.

\subsection{Proposed EmoReg Architecture}
\begin{figure*}[ht!]
    \centering    
    \includegraphics[width=\linewidth]{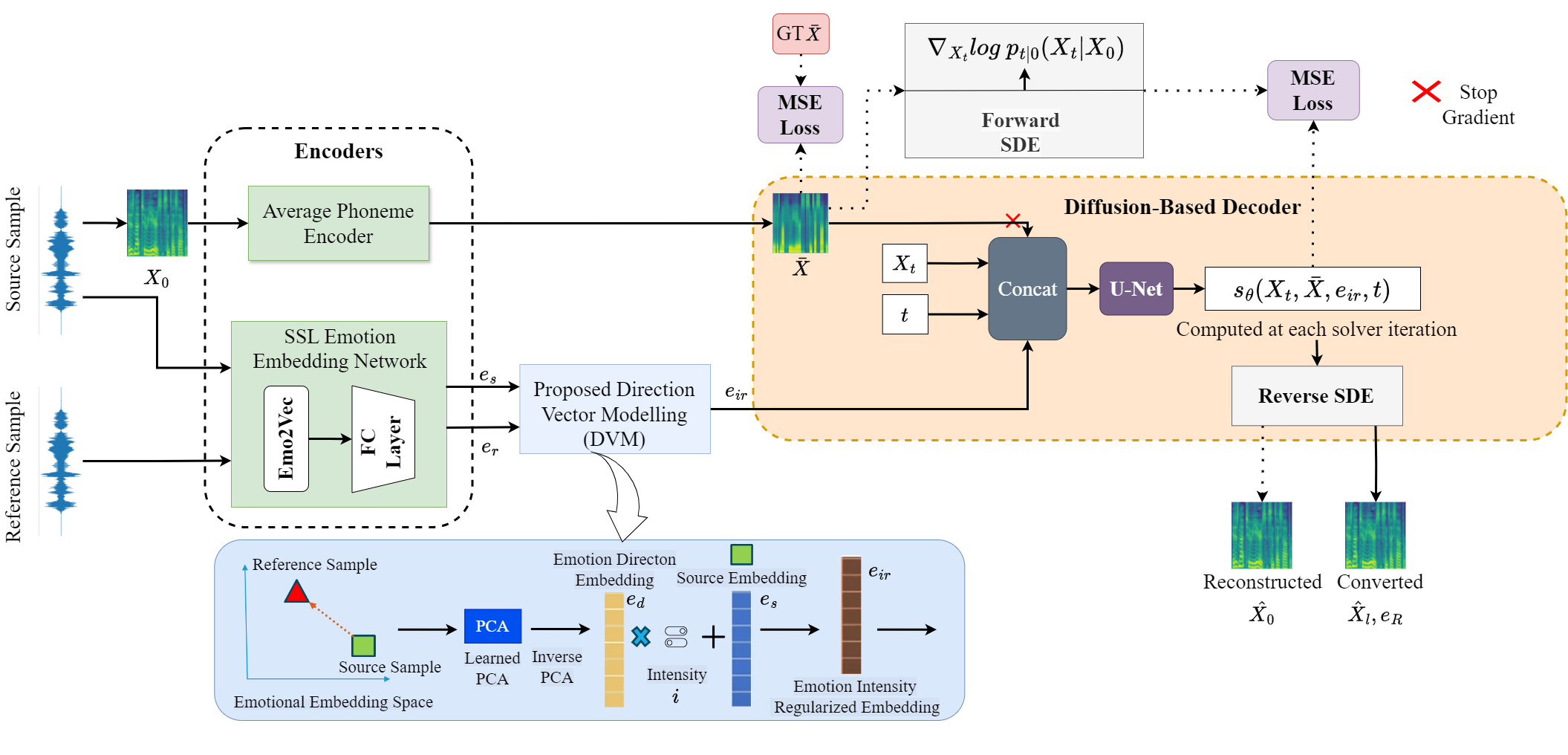}
    \caption{Block diagram of the proposed DVM-based Emotion Intensity Regularized EVC architecture. Dotted arrows represents operations performed only during training. Also, GT $\Bar{X}$ are derived by replacing each phoneme Mel-spectrogram feature in the input with its corresponding pre-calculated average feature. }
    \label{fig:block_diag}
\end{figure*}

The proposed EmoReg architecture uses diffusion-based model to achieve EVC while controlling the emotion intensity $i$. It comprises a diffusion-based decoder and a set of encoders illustrated in Figure \ref{fig:block_diag}. The diffusion decoder is responsible for emotion-controllable speech synthesis, while the encoders individually encode the emotion and speech representation that require disentanglement. The decoder is conditioned on the proposed Direction Vector Modeling (DVM) based features $e_{ir}$, which facilitate the transition between different emotions.  At the output of the diffusion decoder, we obtain the Mel-spectrogram $Y$, which can be converted to the output speech signal $y=H(Y)$, with $H(.)$ representing the HiFiGAN vocoder. Subsequent sections delve deeper into the architectural components of the proposed method.

\subsection{Encoders}
Our EmoReg model consists of two encoders. The first encoder is used to encode the lexical content, while the second one is used to obtain emotion representations.
\subsubsection{Phoneme Encoder:}
In this part, we encode the lexical content at the phoneme level by using speaker- and emotion-independent average phoneme-level Mel characteristics. To achieve this, we utilize the transformer-based encoder as described in \cite{popov2021diffusion}, which has previously been used in speech conversion applications.
\begin{equation}
\label{eq:eq1}
    \overline{\textbf{X}} = \phi(\textbf{X}_0)
\end{equation}
Here, $\overline{\textbf{X}}$ is average Mel-spectrogram of source audio, $\textbf{X}_0$ is source speech Mel-spectrogram and $\phi(.)$ represents the pre-trained average phoneme encoder. 
\subsubsection{SSL Emotion Embedding Network:}
Due to the unavailability of a large annotated emotional speech database, we plan to utilize SSL-based representations. We fine-tune the pre-trained SSL emotion2vec \cite{ma2023emotion2vec} embedding network $E(.)$ using English and Hindi emotional speech databases to learn emotional embedding representations.

We used a pretrained emotion2vec model \cite{ma2023emotion2vec} to generate 768-dimensional embeddings, which we then fine-tuned for classifying target emotions. Similar to the diffusion-based text generation models \cite{gao2022difformer}, we also find that in the high dimensional embedding space, the insufficient noise results in a simple denoising task, which leads to the detoriation of the model. In the process, we first passed these embeddings through a fully connected layer to reduce them to 256 dimensions. We then passed these reduced embeddings through an output layer to classify them into the target emotions (neutral, happy, sad and angry). The training was conducted in a supervised manner. After training, we used the network with the newly added fully connected layer to obtain 256-dimensional emotion embeddings. These emotional embeddings are conditioned in the decoder of the diffusion model in order to achieve target emotional state in the converted output. Additionally, these emotional embeddings are manipulated via proposed directional vector modelling to obtain a fine control over intensities associated with the target emotional states.
\subsection{Direction Vector Modeling}
\begin{figure}[h]
    \centering    
    \includegraphics[width=\linewidth]{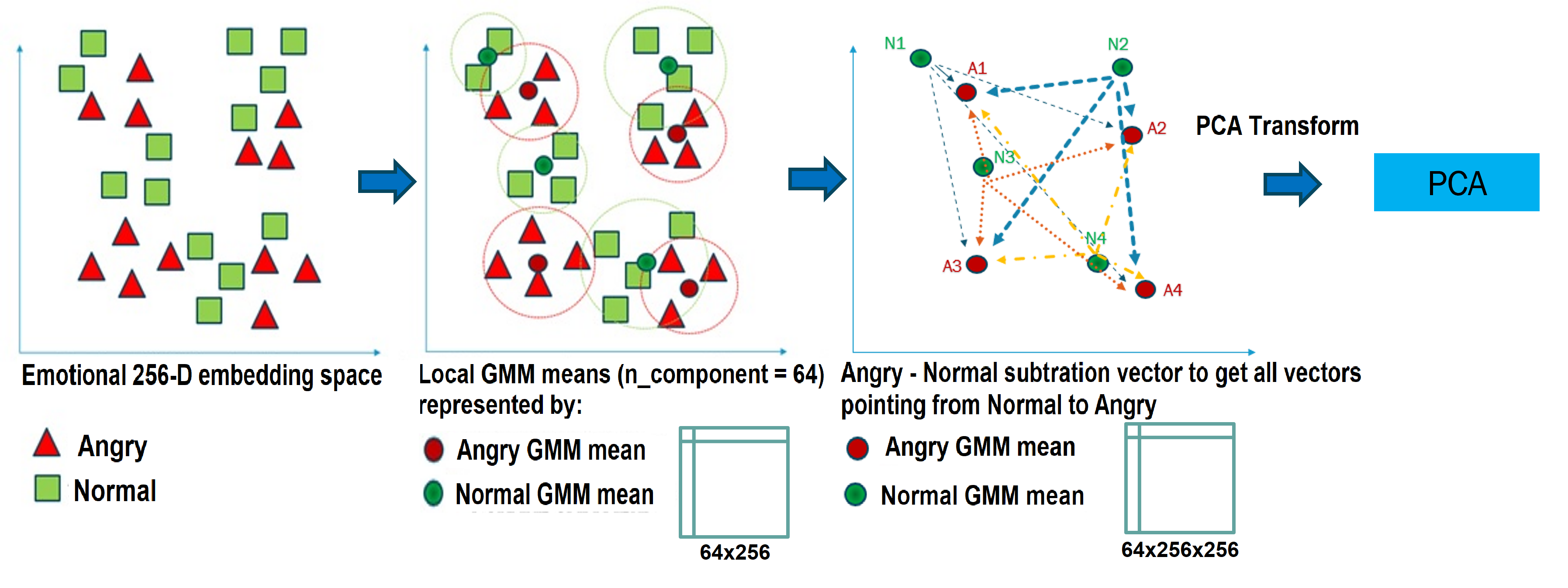}
    \caption{Three key steps of the proposed DVM approach. 1) Fitting local GMM to each emotional state. 2) computing directional vectors for all possible transitions from the local mean of one emotional state to another. 3) Applying PCA to find relevant direction for emotional transition.}
    \label{fig:dvm}
\end{figure}

During training phase, the 256-dimension emotional embeddings from SSL Emotion Embedding Network are used as an input to the DVM module as shown in Figure \ref{fig:dvm}. We modeled the emotion embedding space using a 64-component Gaussian Mixture Model (GMM) \cite{reynolds2009gaussian}  to derive the local mean vector for each emotion, i.e., Angry, Happy, Neutral, and Sad. After extracting the GMM features for each emotion, pairwise subtraction is performed between the embeddings of Angry, Happy, and Sad with those of Neutral to derive the emotional direction vector matrix for all possible transitions from local mean of one emotional state to another. Subsequently, Principal Component Analysis (PCA) \cite{abdi2010principal} is applied to the emotional direction matrix corresponding to each emotion, reducing it to 128 components since not all 256 components are reflecting changes related to emotional states, they may also be affected due to content variability. Ablation analysis for selecting number of principle components is presented in Table 5.

During the inference phase, given a source sample in the Neutral emotion and a reference sample in another emotion (e.g., Angry), the emotional direction vector is calculated by subtracting the source's neutral embedding ($e_s$) from the reference's emotional embedding ($e_r$). This direction vector is then transformed using the PCA to obtain a 128-dimensional vector, which is then converted back to the original 256-dimensional space using PCA inverse transformation, resulting in the final emotional reference embedding $e_d$. The emotional reference embedding is then scaled by a specified factor corresponding to the intensity value $i$, which ranges between 0 to 1, and added to the source embedding (i.e., $e_s$) to shift it towards the reference emotion (i.e., $e_{ir}$) in the embedding space. The diffusion model's reverse Stochastic Differential Equation (SDE) process is conditioned on the resulting scaled embeddings, which effectively transform the source sample into the output speech with a reference target emotion regulated by the emotion intensity value.

\subsection{Diffusion-based Decoder}
The diffusion-based decoder follows the SDE formalism as described in \cite{popov2021grad}. Consider the continuous diffusion time-step variable $t$, which characterizes the progression of the diffusion process. The forward SDE for $0 < t \leq 1$ is given by:
\begin{equation}
\label{eq:eq2}
    d\textbf{X}_{t} = \frac{1}{2}\beta_{t}(\overline{\textbf{X}} - \textbf{X}_{t})dt + \sqrt{\beta_{t}}d\textbf{w}
\end{equation}
where $\textbf{X}_{t}$ represents the current process state with initial condition 
$\textbf{X}_{0}$ 
and $\textbf{w}$ is the typical Wiener process \cite{karatzas2014brownian}. The noise schedule is represented by a non-negative function $\beta_{t}$. The mean evolution here represents an interpolation that ends roughly at the distribution of average voice phoneme characteristics $\overline{\textbf{X}}$ at $t = 1$ and begins at the distribution of source $\textbf{X}_{0}$ at $t = 0$. There is an associated reverse SDE with the forward SDE as given in Eq. \ref{eq:eq2}.
\begin{equation}
    d\textbf{X}_{t} = \Bigr[-\frac{1}{2}\beta_{t}( \overline{\textbf{X}}-\textbf{X}_{t}) + \beta_{t}\nabla_{X_{t}} log p_{t} (\textbf{X}_{t}|\overline{\textbf{X}}) \Bigr] dt + \beta_{t} d\Tilde{\textbf{w}}
\end{equation}
where $d\Tilde{\textbf{w}}$ is the Wiener process which at each diffusion time-step moving backward. Here noise $\beta_{t} = \beta_{0} + t(\beta_{1} - \beta_{0})$ follows linear schedule for both $\beta_{0}$ and $\beta_{1}$ such that $e^{-\int_{0}^{1}(\beta_{s}ds)}$ tends to zero \cite{popov2021diffusion}. Additionally, the reverse SDE has the same trajectory as the forward SDE; that is, it begins roughly with the distribution of average-voice and ends at $t = 0$ with the distribution of source-targets. Here, we employ the score model $s_{\theta}$ as the U-Net architecture from \cite{ronneberger2015u}. The score model takes $\textbf{X}_{t}$, $t$ and regulated emotion intensity features $e_{ir}=f(e_s,e_r,i)$, where $f(.)$ represents function for the DVM approach. Hence, given the emotion intensity regularized embedding $e_{ir}$, we can then utilize the reverse SDE with the learned $s_{\theta}(\textbf{X}_{t}, \overline{\textbf{X}}, t, e_{ir})$ to estimate the reconstructed $\textbf{X}_{0}$ from the average voice $\overline{\textbf{X}}$.

While doing inference, a source sample in a neutral emotion is converted to a specified target emotion by utilizing a reference sample in the corresponding target emotion. Both the source emotional embedding and the reference emotional embedding are given as input to the DVM, which models the direction vector and produces an emotion intensity regularized embedding. This embedding and the learned score function are then used to transform the source sample into the reference sample emotion.

\section{Experimental Analysis}
\subsection{Dataset}
Our performance evaluation primarily uses the Emotional Speech Database (ESD) \cite{zhou2021seen}. The ESD dataset contains 29 hours of 350 parallel recorded utterances in five distinct emotions, i.e., Neutral, Happy, Angry, Sad, and Surprise, from ten English speakers, respectively. In addition, we utilized 36 hours of internally developed non-parallel Hindi emotional speech database recorded by 9 native Hindi speakers for evaluating performance across languages because emotional speech database is not publicly available for other Indian languages. To record this database, we have collected Hindi texts from stories and categorized these texts into four emotions, namely, Neutral, Happy, Angry, and Sad. Each speaker has recorded one hour of data on each emotion. We used a 90:10 train-test split for both ESD-English (excluding Chinese) and Hindi data. We presented our analysis for three emotional state conversion scenarios along with intensity control: neutral-to-happy, neutral-to-sad, and neutral-to-angry. Note, we use nonparallel reference emotional samples from different speaker for evaluation.
\subsection{Implementation Details}
Initially, emotion-independent speech representation is achieved by using average phoneme-level Mel-spectrogram characteristics. The process begins with aligning speech frames with phonemes in the ESD dataset using the Montreal Forced Aligner (MFA) \cite{mcauliffe2017montreal}. Once aligned, the Mel features of each phoneme are aggregated across the entire dataset to obtain their average Mel-spectrogram features. The Phoneme Encoder is then trained to minimize the mean square error between the output Mel-spectrograms and the ground truth average voice Mel-spectrograms. These ground truth spectrograms are derived by replacing each phoneme Mel-spectrogram feature in the input with its corresponding average feature calculated earlier. The encoder is trained for 300 epochs with a batch size of 32 and a base learning rate of 5e-5 using the Adam optimizer. Our decoder is built on top of DiffVC architecture \cite{popov2021diffusion}. The number of parameters in the Encoder is 8.5mn and the decoder is 118mn. The decoder is trained for 126 epochs with a batch size of 32 using the Adam optimizer with a base learning rate of 1e-4. The noise schedule parameters $\beta_0$ and $\beta_1$ are set to 0.05 and 20.0, respectively. The model mainly operates on Mel-spectrograms with 80 Mel features and a sampling rate of 22.05 kHz. The entire training process takes approximately 4 hours for the encoder and 12 hours for the decoder, both trained on an NVIDIA L40S GPU. 

\subsection{Baseline Methods}
\begin{itemize}
    \item \textbf{EmoVox \cite{zhou2022emotion}}: This approach explicitly controls the emotion intensity in the EVC task by using characteristic features to express fine-grained emotion intensity and further learn the actual emotion encoder from an emotion-labeled database.
    \item \textbf{Mixed Emotion \cite{zhou2022mixed}}: It is a two stages seq-to-seq training method for emotional voice conversion. It uses a multi-speaker TTS corpus to perform style initialization in stage 1 to separate language content from speaking style. Stage 2 uses a limited amount of emotional speech data for emotion training to learn how to disentangle the speech's linguistic and emotional style information. 
    \item \textbf{Other baselines}: We have also considered CycleGAN-EVC \cite{fu2021cycletransgan}, StarGAN-EVC \cite{rizos2020stargan}, StyleVC \cite{du2021disentanglement}, DISSC \cite{maimon2022speaking} and Seq2Seq-EVC \cite{zhou2021limited} to compare performance of proposed EmoReg approach w.r.t emotion voice conversion. However, we have excluded them from intensity control related evaluations as these approaches do not support emotion intensity regularization.
    \item \textbf{Ablation EmoReg w/o DVM}: Here, we utilized the proposed EmoReg architecture without directional vector modeling. In particular, we applied to scale in the direction vector learned between the global mean of target emotion embedding reference and source emotion embedding along with intensity-related scaling function, i.e., via traditional interpolation-based strategy.
\end{itemize}

\subsection{Experimental Results}
\subsubsection{Objective Evaluation}
\begin{figure}[h!]
    \centering      
    \includegraphics[width=\linewidth]{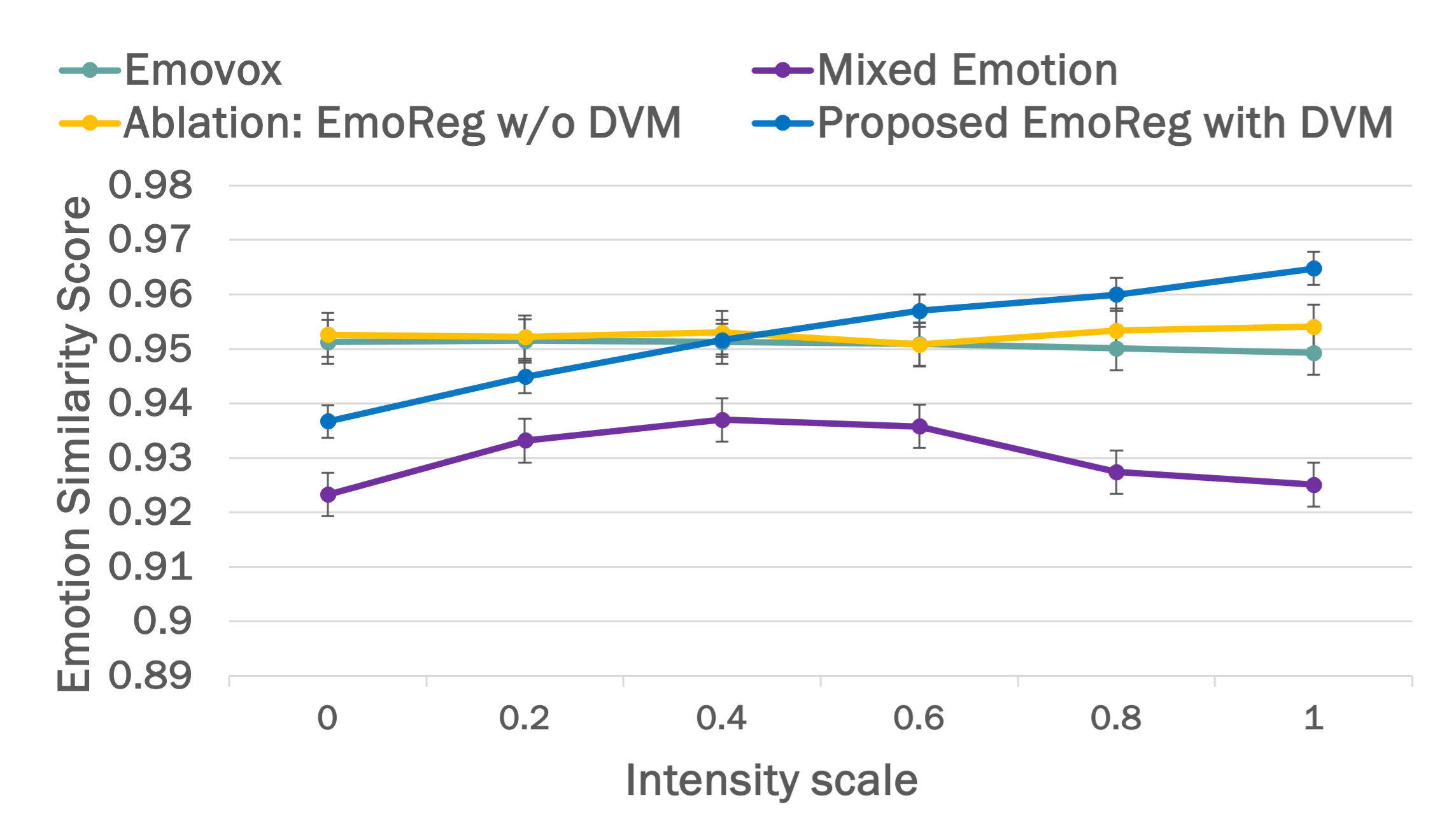}
    \caption{Analysis of emotion similarity score with respect to incremental emotion intensity scale.}
    \label{fig:emo_int}
\end{figure}

Since our key goal is obtaining control over the emotion intensity, we primarily considered the emotion similarity score as an objective evaluation metric. We use it to measure the effectiveness of the proposed approach for the emotion conversion and intensity regularization tasks. It is obtained by computing the cosine distance between the generated and ground truth speech emotion embeddings from the pre-trained emotion classifiers. Furthermore, Word Error Rate (WER) and Character Error Rate (CER) are considered to measure the intelligibility of the output while performing emotion intensity regularization on a continuous scale. Here, we used Whisper-small model \cite{radford2023robust}. The emotion similarity score for emotion voice conversion is calculated for Neutral-to-Angry, Neutral-to-Sad, and Neutral-to-Happy emotion conversion scenarios for the proposed approach and baseline methods, as shown in Table \ref{tab:sota_comp}. These are denoted by the abbreviations: Neu-Ang, Neu-Sad, and Neu-Hap, respectively. Similarly, we achieved improvements of 2 to 11\% with the proposed DVM-based approach in AutoPCP-based objective evaluations for emotional similarity \cite{barrault2023seamless} and scale-wise subjective evaluations for emotional controllability. Detail results are omitted due to space constraints.
\begin{table}[h!]
\caption{Analysis of emotion similarity scores along with margin of error corresponding to the 95\% CI.}
\label{tab:sota_comp}
\resizebox{\columnwidth}{!}{%
\begin{tabular}{ccccc}
\hline
\textbf{Methods} & \textbf{Neu-Ang $\uparrow$}                      & \textbf{Neu-Sad $\uparrow$}      & \textbf{Neu-Hap $\uparrow$}                    & \textbf{Average$\uparrow$}                     \\ \hline
Emovox           & 0.94 $\pm$ 0.004                     & 0.94 $\pm$ 0.004                     & 0.95 $\pm$ 0.004                     & 0.94 $\pm$ 0.004                     \\
Mixed Emotion    & 0.94 $\pm$ 0.004                     & 0.92 $\pm$ 0.004                     & 0.90 $\pm$ 0.004                     & 0.92 $\pm$ 0.004                     \\
CycleGAN-EVC     & 0.96 $\pm$ 0.004                     & 0.92 $\pm$ 0.004                     & 0.91 $\pm$ 0.004                     & 0.93 $\pm$ 0.004                     \\
StarGAN-EVC      & 0.95 $\pm$ 0.004                     & 0.91 $\pm$ 0.004                     & 0.91 $\pm$ 0.004                     & 0.93 $\pm$ 0.004                     \\
Seq2Seq-EVC      & 0.96 $\pm$ 0.004                     & 0.93 $\pm$ 0.004                     & 0.87 $\pm$ 0.004                     & 0.92 $\pm$ 0.004                     \\
StyleVC          & \multicolumn{1}{l}{0.96 $\pm$ 0.004} & \multicolumn{1}{l}{0.92$\pm$ 0.004}  & \multicolumn{1}{l}{0.91$\pm$ 0.004}  & \multicolumn{1}{l}{0.93 $\pm$ 0.004} \\
DISSC            & \multicolumn{1}{l}{0.88 $\pm$ 0.004} & \multicolumn{1}{l}{0.91 $\pm$ 0.004} & \multicolumn{1}{l}{0.87 $\pm$ 0.004} & \multicolumn{1}{l}{0.89 $\pm$ 0.004} \\
Ablation         & 0.96 $\pm$ 0.004                     & 0.93 $\pm$ 0.004                     & 0.95 $\pm$ 0.004                     & 0.94 $\pm$ 0.004                     \\
Proposed         & \textbf{0.97 $\pm$ 0.003}            & \textbf{0.96 $\pm$ 0.003}            & \textbf{0.95 $\pm$ 0.003}            & \textbf{0.96 $\pm$ 0.003}            \\ \hline
\end{tabular}
}
\end{table}
Table \ref{tab:sota_comp} illustrates the comprehensive evaluation of the proposed EmoReg with DVM, which outperforms all the SOTA approaches in the EVC task. The performance of the EmoReg approach is further investigated on different intensity scores using emotion similarity score, as shown in Figure \ref{fig:emo_int}. We used an intensity scale range of 0 to 1 with a 0.2 step size and we could only consider EmoVox \cite{zhou2022emotion} and MixedEmotion \cite{zhou2022mixed} models, as others i.e., CycleGAN-StarGAN-EVC and Seq2Seq-EVC lack support for emotion intensity regularization due to their architectural limitations.
\label{sec:ExperimentalSetup}

Additionally, from Figure \ref{fig:emo_int}, it is apparent that the emotion similarity score increases with an increase in emotion intensity scale which shows that the proposed EmoReg with DVM can achieve fine control over emotion intensity while doing EVC. Whereas, the emotion similarity score of the other baselines and ablation does not vary with an increase in intensity scale and hence, fails to achieve fine control over emotion intensity.

To evaluate the intelligibility of EVC-generated speech, we calculated the WER and CER for baselines and proposed EmoReg. From Table \ref{tab:wer_cer}, we can see that our approach achieves lower WER for the considered intensity scales and, on average, compared to EmovoX and Mixed Emotion models. It further signifies that the proposed EmoReg effectively maintains speech intelligibility during EVC while allowing fine control over emotion intensity. Interestingly, the ablation model of EmoReg without DVM also achieves lower WER at intensity scales of 0.0, 0.2, and 0.4. This could be due to the fact that, within a small range of emotional intensity, traversing within a closely related emotional space often leads to similar emotional state representations. However, our approach with DVM estimates the optimum direction for transitioning from one emotional state to another. Hence, after 0.6 intensity-related distance, the model achieves higher emotional similarity scores and lower WER/CER scores.    
\begin{table}[h!]
\caption{Comparison of WER and CER for proposed approach with baseline methods for English language.}
\label{tab:wer_cer}
\resizebox{\columnwidth}{!}{%
\begin{tabular}{cccccccc}
\hline
\multicolumn{1}{c}{\textbf{Methods}}                  & \multicolumn{1}{c}{\textbf{0}} & \multicolumn{1}{c}{\textbf{0.2}} & \multicolumn{1}{c}{\textbf{0.4}} & \multicolumn{1}{c}{\textbf{0.6}} & \multicolumn{1}{c}{\textbf{0.8}} & \multicolumn{1}{c}{\textbf{1}} & \textbf{Avg}               \\ \hline
\multicolumn{8}{c}{\textbf{WER $\downarrow$}}                                                                                                                                                                                                                                                                                                                                                                                                                                                        \\ \hline
\multicolumn{1}{c}{Emovox}              & \multicolumn{1}{c}{34.82}                              & \multicolumn{1}{c}{34.52}                                & \multicolumn{1}{c}{34.43}                                & \multicolumn{1}{c}{35.86}                                & \multicolumn{1}{c}{39.15}                                & \multicolumn{1}{c}{38.57}                              & 36.23                      \\ 
\multicolumn{1}{c}{Mixed Emotion}         & \multicolumn{1}{c}{72.22}                              & \multicolumn{1}{c}{55.56}                                & \multicolumn{1}{c}{75.89}                                & \multicolumn{1}{c}{70}                                   & \multicolumn{1}{c}{71.12}                                & \multicolumn{1}{c}{77.78}                              & 70.43                      \\ 
\multicolumn{1}{c}{Ablation}          & \multicolumn{1}{c}{\textbf{13.86}}                     & \multicolumn{1}{c}{\textbf{13.34}}                       & \multicolumn{1}{c}{\textbf{13.75}}                       & \multicolumn{1}{c}{14.28}                                & \multicolumn{1}{c}{13.76}                                & \multicolumn{1}{c}{14.29}                              & 13.88                      \\ 
\multicolumn{1}{c}{\textbf{Proposed}} & \multicolumn{1}{c}{14.6}                               & \multicolumn{1}{c}{13.65}                                & \multicolumn{1}{c}{13.89}                                & \multicolumn{1}{c}{\textbf{13.09}}                       & \multicolumn{1}{c}{\textbf{13.25}}                       & \multicolumn{1}{c}{\textbf{13.65}}                     & \textbf{13.69}             \\ \hline
\multicolumn{8}{c}{\textbf{CER $\downarrow$}}                                                                                                                                                                                                                                                                                                                                                                                                                                                        \\ \hline
\multicolumn{1}{c}{Emovox}              & \multicolumn{1}{l}{21.82}                              & \multicolumn{1}{l}{20.57}                                & \multicolumn{1}{l}{22.72}                                & \multicolumn{1}{l}{21.63}                                & \multicolumn{1}{l}{24.5}                                 & \multicolumn{1}{l}{23.49}                              & \multicolumn{1}{l}{22.46} \\ 
\multicolumn{1}{c}{Mixed Emotion}         & \multicolumn{1}{l}{36.57}                              & \multicolumn{1}{l}{30.1}                                 & \multicolumn{1}{l}{38.65}                                & \multicolumn{1}{l}{36.11}                                & \multicolumn{1}{l}{38.89}                                & \multicolumn{1}{l}{43.06}                              & \multicolumn{1}{l}{37.23} \\ 
\multicolumn{1}{c}{Ablation}          & \multicolumn{1}{l}{\textbf{8.27}}                               & \multicolumn{1}{l}{8.03}                                 & \multicolumn{1}{l}{8.18}                                 & \multicolumn{1}{l}{7.94}                                 & \multicolumn{1}{l}{8.18}                                 & \multicolumn{1}{l}{8.12}                               & \multicolumn{1}{l}{8.12}  \\ 
\multicolumn{1}{c}{\textbf{Proposed}} & \multicolumn{1}{l}{8.7}                                & \multicolumn{1}{l}{\textbf{7.98}}                                 & \multicolumn{1}{l}{\textbf{8.03}}                                 & \multicolumn{1}{l}{\textbf{7.78}}                                 & \multicolumn{1}{l}{\textbf{7.93}}                                 & \multicolumn{1}{l}{\textbf{8.02}}                               & \multicolumn{1}{l}{\textbf{8.07}}  \\ \hline
\end{tabular}%
}
\end{table}
\begin{table}[h]
\caption{Emotion Similarity scores across languages along with 95 \% confidence interval.}
\label{tab:esc_acr_lang}
\resizebox{\columnwidth}{!}{%
\begin{tabular}{ccccc}
\hline
\multicolumn{1}{c}{\textbf{Methods}} & \multicolumn{1}{c}{\textbf{Neu-Ang $\uparrow$}}          & \multicolumn{1}{c}{\textbf{Neu-Sad $\uparrow$}}          & \multicolumn{1}{c}{\textbf{Neu-Hap $\uparrow$}}          & \textbf{Avg $\uparrow$}              \\ \hline
\multicolumn{5}{c}{\textbf{English}}                                                                                                                                                                                         \\ \hline
\multicolumn{1}{c}{Ablation}         & \multicolumn{1}{c}{0.96 $\pm$ 0.004}          & \multicolumn{1}{c}{0.93 $\pm$ 0.004}          & \multicolumn{1}{c}{0.95 $\pm$ 0.004}          & 0.94 $\pm$ 0.004          \\ 
\multicolumn{1}{c}{Proposed}         & \multicolumn{1}{c}{\textbf{0.97 $\pm$ 0.003}} & \multicolumn{1}{c}{\textbf{0.96 $\pm$ 0.003}} & \multicolumn{1}{c}{\textbf{0.95 $\pm$ 0.003}} & \textbf{0.96 $\pm$ 0.003} \\ \hline
\multicolumn{5}{c}{\textbf{Hindi}}                                                                                                                                                                                           \\ \hline
\multicolumn{1}{c}{Ablation}         & \multicolumn{1}{c}{0.89 $\pm$ 0.003} & \multicolumn{1}{c}{0.86 $\pm$ 0.003} & \multicolumn{1}{c}{\textbf{0.89 $\pm$ 0.003}} & \textbf{0.88 $\pm$ 0.003} \\ 
\multicolumn{1}{c}{Proposed}         & \multicolumn{1}{c}{\textbf{0.91 $\pm$ 0.003}} & \multicolumn{1}{c}{\textbf{0.87 $\pm$ 0.003}} & \multicolumn{1}{c|}{0.87 $\pm$ 0.003}          & \textbf{0.88 $\pm$ 0.003} \\ \hline
\end{tabular}
}
\end{table}
\begin{figure}[h!]
    \centering    
    \includegraphics[width=\linewidth]{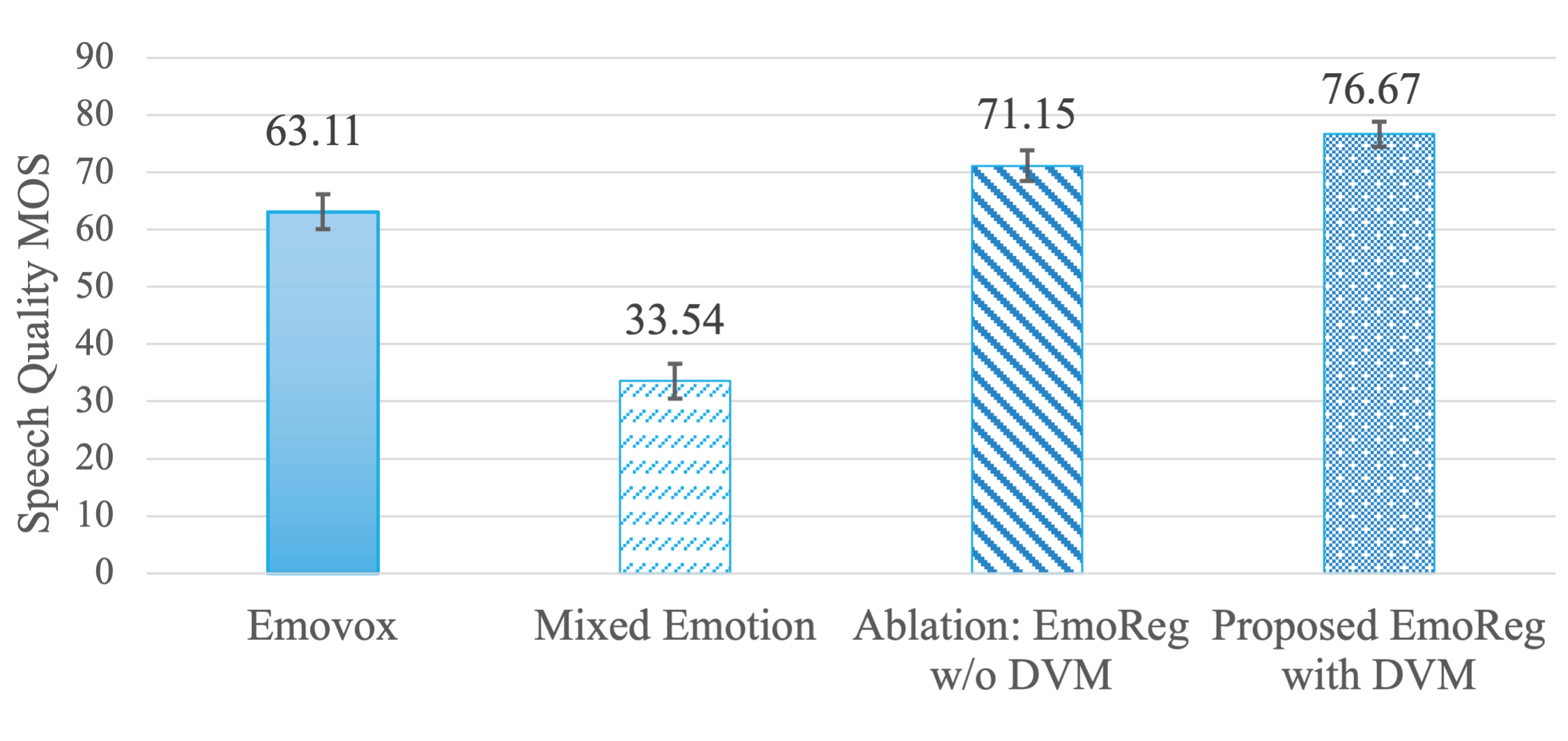}
    \caption{MUSHRA-based MOS scores for speech quality for proposed EmoReg approach and baseline methods.}
    \label{fig:mos_esd}
\end{figure}
\begin{figure*}[h]
    \centering    
    \includegraphics[width=\linewidth]{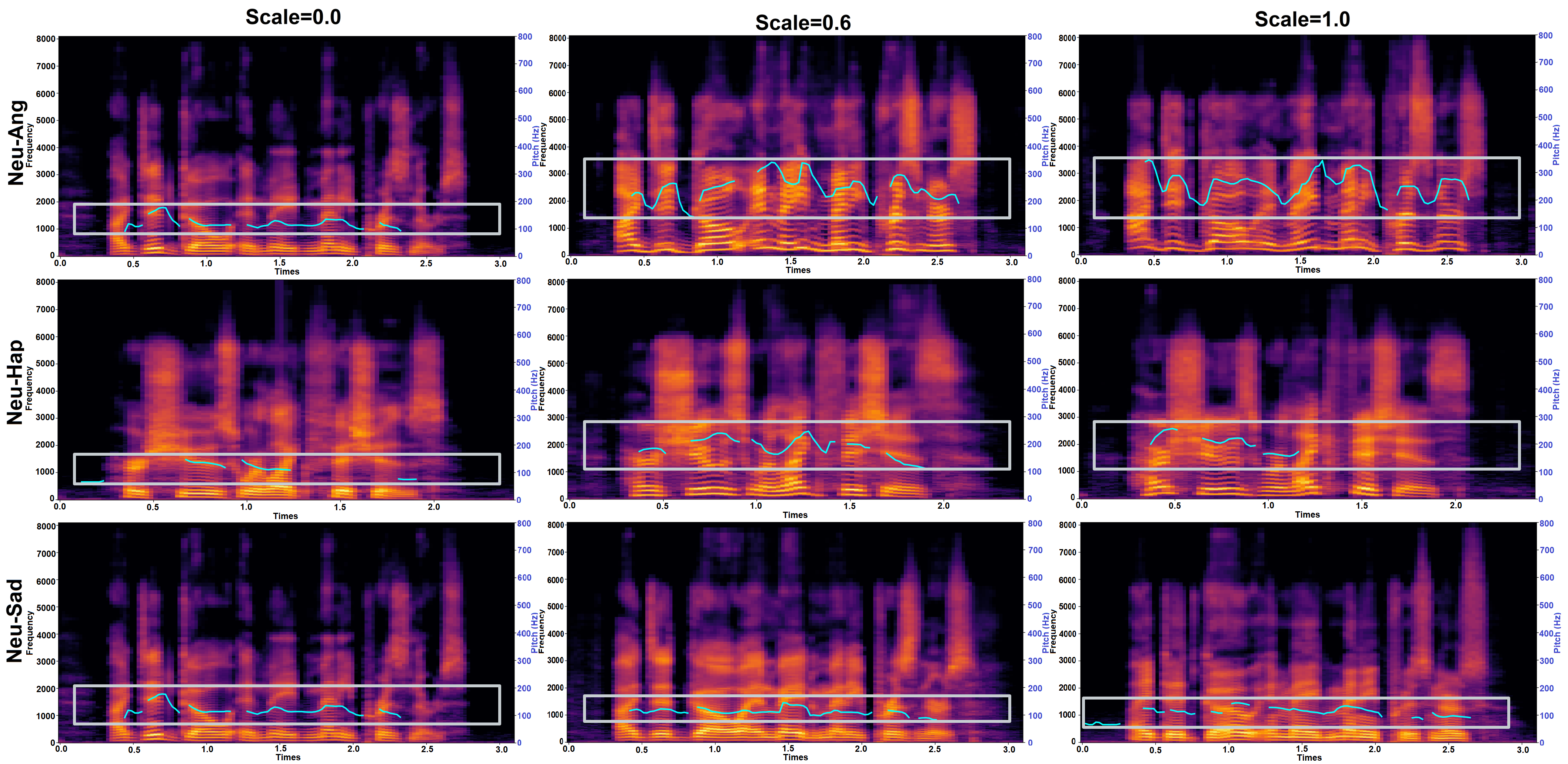}
    \caption{Visualization of Mel-spectrogram and pitch variation for emotion conversion from Neutral to Angry, Happy and Sad emotion with different intensity scale.}
    \label{fig:spect_analys}
\end{figure*}
\subsubsection{Subjective Evaluation}
For the subjective evaluation, we considered the Mean Opinion Score (MOS) metric on overall speech quality. A total of 25 individuals with no known hearing impairment between 25 and 35 years of age participated in the subjective evaluation. We asked subjects to rate each sample based on quality on a scale of 0 to 100, where 100 indicates the highest and 0 lowest quality. The evaluation of the proposed EmoReg, compared to baseline methods on the ESD dataset, is depicted in Figure \ref{fig:mos_esd}. The figure shows that the proposed approach yields an absolute increment of 13.56\% and 43.13\% in the MOS value for quality compared to the baselines EmoVox and Mixed Emotion, respectively. This noticeable quality gap is evident in the samples we provided and samples from the baselines. Diffusion model-based architectures are one key reason for achieving high-quality output. Moreover, the proposed EmoReg architecture with DVM achieves a 5.52\% absolute increment in MOS compared to the ablation model without DVM. This indicates that the proposed DVM is helping in traversing emotion embedding space with improved-quality samples. In addition, We also obtained a similar improvement in the performance (1\%) with the proposed DVM in the non-diffusion-based architecture, namely VITS. Detail results are omitted due to space constraint.   



\subsubsection{Performance Across Languages:}
The effectiveness of the proposed approach is evaluated across different databases using similar objective and subjective assessments for both English and Hindi languages. Table \ref{tab:esc_acr_lang} shows the emotion similarity scores for Neutral-to-Angry, Neutral-to-Sad, and Neutral-to-Happy emotion voice conversion for ablation and the proposed EmoReg approach for both languages. It is evident from Table \ref{tab:esc_acr_lang} that the proposed approach also performs well for the Hindi language. Additionally, the intelligibility of converted speech in both languages is evaluated using WER and CER along with MOS score and is presented in Table \ref{tab:acr_lang}.

\begin{table}[h!]
\caption{Comparison of WER, CER and MUSHRA based MOS scores for the ablation and the proposed approaches.}
\label{tab:acr_lang}
\resizebox{\columnwidth}{!}{%
\begin{tabular}{ccccccc}
\hline
\multirow{2}{*}{\textbf{Methods}}                                   & \multicolumn{2}{c}{\textbf{WER $\downarrow$}}  & \multicolumn{2}{c}{\textbf{CER $\downarrow$}}  & \multicolumn{2}{c}{\textbf{MOS $\uparrow$}}  \\ \cline{2-7} 
& \textbf{English} & \textbf{Hindi} & \textbf{English} & \textbf{Hindi} & \textbf{English} & \textbf{Hindi} \\ \hline
\begin{tabular}[c]{@{}c@{}}Ablation\end{tabular} & 13.88            & 33.33          & 8.12             & \textbf{17.74} & 71.15            & 80.07          \\
\begin{tabular}[c]{@{}c@{}}Proposed \end{tabular} & \textbf{13.69}   & \textbf{32.93} & \textbf{8.07}    & 18.20          & \textbf{76.67}   & \textbf{80.58} \\ \hline
\end{tabular}%
}
\end{table}

\begin{table}[h!]
\centering
\caption{Ablation analysis of emotional similarity score for different number of
PCA components.}
\label{tab:pca_conmponents}
\resizebox{0.9\columnwidth}{!}{%
\begin{tabular}{ccccccc}
\hline
\multirow{2}{*}{\textbf{Methods}} & \multicolumn{3}{c}{\textbf{English}}                                                 & \multicolumn{3}{c}{\textbf{Hindi}}                                                     \\ \cline{2-7} 
                                  & \multicolumn{1}{c}{\textbf{64}} & \multicolumn{1}{c}{\textbf{128}}   & \textbf{256} & \multicolumn{1}{c}{\textbf{64}} & \multicolumn{1}{c}{\textbf{128}}   & \textbf{256}   \\ \hline
\textbf{Neu-Ang $\uparrow$}                  & \multicolumn{1}{c}{\textbf{0.938}}       & \multicolumn{1}{c}{\textbf{0.938}} & 0.937        & \multicolumn{1}{c}{0.889}       & \multicolumn{1}{c}{\textbf{0.898}} & 0.896          \\ 
\textbf{Neu-Sad $\uparrow$}                  & \multicolumn{1}{c}{0.949}       & \multicolumn{1}{c}{\textbf{0.950}} & 0.944        & \multicolumn{1}{c}{0.889}       & \multicolumn{1}{c}{0.889}          & \textbf{0.896} \\ 
\textbf{Neu-Hap $\uparrow$}                  & \multicolumn{1}{c}{0.911}       & \multicolumn{1}{c}{\textbf{0.934}} & 0.928        & \multicolumn{1}{c}{0.849}       & \multicolumn{1}{c}{\textbf{0.858}} & 0.851          \\ \hline
\end{tabular}%
}
\end{table}
We conducted an ablation analysis of emotional similarity scores using different numbers of PCA components for both English and Hindi languages. PCA is primarily applied to the direction vectors obtained among all possible combinations of transitions from local mean vectors presented in two different emotional states. While these direction vectors primarily reflect changes related to emotional states, they may also be influenced by content and gender differences, as emotional representations are not entirely independent of language or gender. To identify direction vectors corresponding to emotional state differences, we selected three different numbers of principal components (64, 128, and 256) to capture variability, focusing primarily on emotional state transitions. As shown in Table \ref{tab:pca_conmponents}, 128 components yield the best emotion similarity scores for both languages. Therefore, 128 principal components were chosen for the proposed DVM-based approach.
\subsection{Visual Analysis}
Additionally, we present the variations in prosody patterns, such as pitch and Mel-spectrogram, for various emotions, Neutral-to-Angry, Neutra-to-Sad, and Neutral-to-Happy in Figure \ref{fig:spect_analys}. For Neutral-to-Angry emotion, we can see that with an increase in emotion intensity scale, pitch frequency range and dynamics also increase which is indicated by a white box. As harmonics are an integer multiple of the fundamental frequency, we can also see that for scales 0.6 and 1.0, the harmonics are wider than those on scale 0. It indicates that the proposed approach can control the intensity of emotion. Similar observations are seen in the case of Neutral-to-Happy emotion, where the pitch range increases with an increase in intensity scale. For the Neutral-to-Sad condition, the harmonics in the spectrogram are getting narrower with an increase in intensity scale because sadness is a low pitch-flattish emotion where the pitch range is usually lower as compared to high pitch with high variance emotions like anger and happiness.
\section{Conclusions}
In this paper, we introduced the EmoReg model for emotion voice conversion with emotion intensity regularization. By leveraging SSL-based emotion embeddings, we achieved effective emotion representation from speech. We proposed a direction vector modeling (DVM) to transition between emotional states while controlling emotion intensity. We evaluated our approach against the SOTA architectures for both English and Hindi languages. Our approach demonstrated significant improvements of 13.56\% over EmoVox and 43.13\% over Mixed Emotion in speech quality. Additionally, the proposed EmoReg model outperformed existing methods in various objective evaluations. Moving forward, we aim to extend this work to enhance emotional intensity regularization in challenging environments, such as those with background music or noise.  
\label{sec:SummaryConclusion}
\bibliography{aaai25}
\appendix
\section{Appendix}
\subsection{Additional Subjective and Objective Evaluations}
Due to space constraints in the original manuscript, we are adding additional subjective and objective evaluations in this Appendix. 
In particular, we conducted a subjective evaluation for the proposed model at different intensity scales as shown in Table 6. 
\begin{table}[h]
\caption{Scalewise subjective evaluation scores across relevant baselines.}
\resizebox{\columnwidth}{!}{%
\begin{tabular}{ccccccc}
\hline
Methods       & 0.0    & 0.2    & 0.4    & 0.6    & 0.8    & 1.0    \\ \hline
Emovox        & 45.846 & 46.949 & 45.590 & 43.949 & 43.513 & 44.000 \\
Mixed Emotion & 34.56  & 42.06  & 47.39  & 51.70  & 51.26  & 54.87  \\
Ablation      & 68.205 & 66.949 & 68.385 & 71.923 & 71.179 & 73.000 \\
Proposed      & 70.000 & 71.923 & 71.974 & 72.256 & 73.026 & 73.103 \\ \hline
\end{tabular}
}
\end{table}
In addition, We have added AutoPCP evaluation scores as shown in Table 7. AutoPCP is an utterance-level estimator used to measure the similarity between two speech samples in terms of prosody. A higher AutoPCP score indicates improved similarity in prosody. We found that the proposed approach achieves 2\%-11\% relative improvements in the AutoPCP scores compared to the SOTA. \\
\begin{table}[h]
\caption{Analysis of AutoPCP-based emotion similarity scores across different SOTA algorithms.}
\resizebox{\columnwidth}{!}{%
\begin{tabular}{ccccc}
\hline
Methods       & Neutral-Angry  & Neutral-Sad     & Neutral-Happy   & Avg             \\ \hline
Emovox        & 3.3555          & 2.9537          & 2.8372          & 3.0488          \\
Mixed Emotion & 3.2550          & 2.7127          & 2.9570          & 2.9749          \\
CycleGAN-EVC  & 3.2115          & 2.5084          & 2.8831          & 2.8677          \\
StarGAN-EVC   & 3.2912          & 2.4995          & 2.7613          & 2.8506          \\
Seq2Seq-EVC   & 3.1873          & 2.8387          & \textbf{2.9844} & 3.0035          \\
StyleVC       & 3.1447          & 2.8557          & 2.7193          & 2.9112          \\
DISSC         & -               & 3.04            & 2.47            & 2.75            \\
Ablation      & 3.3016          & 2.8547          & 2.9115          & 3.0226          \\
Proposed      & \textbf{3.3991} & \textbf{2.9865} & 2.9382          & \textbf{3.1079} \\ \hline
\end{tabular}
}
\end{table}
\indent Moreover, in manuscript, we have only shown results from neutral emotions to non-neutral emotions. However, we have observed that our method works equally well in non-neutral to neutral emotion conversion scenarios as well. 

\end{document}